\documentstyle [12pt] {article}

\catcode`\@=11
\def\@citex[#1]#2{%
\if@filesw \immediate \write \@auxout {\string \citation {#2}}\fi
\@tempcntb\m@ne \let\@h@ld\relax \def\@citea{}%
\@cite{%
  \@for \@citeb:=#2\do {%
    \@ifundefined {b@\@citeb}%
      {\@h@ld\@citea\@tempcntb\m@ne{\bf ?}%
      \@warning {Citation `\@citeb ' on page \thepage \space
undefined}}%
      {\@tempcnta\@tempcntb \advance\@tempcnta\@ne%
      \@tempcntb\number\csname b@\@citeb \endcsname \relax%
      \ifnum\@tempcnta=\@tempcntb 
	\ifx\@h@ld\relax%
	  \edef \@h@ld{\@citea\csname b@\@citeb\endcsname}%
	\else%
	  \edef\@h@ld{\ifmmode{-}\else--\fi\csname
b@\@citeb\endcsname}%
	\fi%
      \else
	\@h@ld\@citea\csname b@\@citeb \endcsname%
	\let\@h@ld\relax%
      \fi}%
    \def\@citea{,\penalty\@highpenalty\,}%
  }\@h@ld
}{#1}}

\def\@citeb#1#2{{[#1]\if@tempswa , #2\fi}}
\def\@citeu#1#2{{$^{#1}$\if@tempswa , #2\fi }}
\def\@citep#1#2{{#1\if@tempswa , #2\fi}}

\def\bcites{         
	\catcode`\@=11
	\let\@cite=\@citeb
	\catcode`\@=12
}

\def\upcites{         
	\catcode`\@=11
	\let\@cite=\@citeu
	\catcode`\@=12
}

\def\plaincites{      
	\catcode`\@=11
	\let\@cite=\@citep
	\catcode`\@=12
}

\newcount\hour
\newcount\minute
\newtoks\amorpm
\hour=\time\divide\hour by 60
\minute=\time{\multiply\hour by 60 \global\advance\minute by-\hour}
\edef\standardtime{{\ifnum\hour<12 \global\amorpm={am}%
	\else\global\amorpm={pm}\advance\hour by-12 \fi
	\ifnum\hour=0 \hour=12 \fi
	\number\hour:\ifnum\minute<10
0\fi\number\minute\the\amorpm}}
\edef\militarytime{\number\hour:\ifnum\minute<10
0\fi\number\minute}

\def\draftlabel#1{{\@bsphack\if@filesw {\let\thepage\relax
   \xdef\@gtempa{\write\@auxout{\string
      \newlabel{#1}{{\@currentlabel}{\thepage}}}}}\@gtempa
   \if@nobreak \ifvmode\nobreak\fi\fi\fi\@esphack}
	\gdef\@eqnlabel{#1}}
\def\@eqnlabel{}
\def\@vacuum{}
\def\marginnote#1{}
\def\draftmarginnote#1{\marginpar{\raggedright\scriptsize\tt#1}}
\def\draft{
	\pagestyle{plain}
	\overfullrule=2pt
	\oddsidemargin -.5truein
	\def\@oddhead{\sl \phantom{\today\quad\militarytime} \hfil
	\smash{\Large\sl DRAFT} \hfil \today\quad\militarytime}
	\let\@evenhead\@oddhead
	\let\label=\draftlabel
	\let\marginnote=\draftmarginnote
	\def\ps@empty{\let\@mkboth\@gobbletwo
	\def\@oddfoot{\hfil \smash{\Large\sl DRAFT} \hfil}
	\let\@evenfoot\@oddhead}

\def\@eqnnum{(\theequation)\rlap{\kern\marginparsep\tt\@eqnlabel}%
	\global\let\@eqnlabel\@vacuum}  }

\def\blackfonts{
	\font\blackboard=msbm10 scaled\magstep1
	\font\blackboards=msbm8
	\font\blackboardss=msbm6
}

\def\nblack{            
	\def\ZZ{{Z \n{10} Z}}
	\def\NN{{N \n{14} N}}
	\def\CC{{C \n{11} C}}
	\def\RR{{R \n{11} R}}
	\def\QQ{{Q \n{12} Q}}
	\def\PP{{P \n{11} P}}
}

\def\prep{         
	\catcode`\@=11
	\input art10.sty
	\catcode`\@=12
	
	\let\small\null
	\def\blackfonts{
		\font\blackboard=msbm10
		\font\blackboards=msbm7
		\font\blackboardss=msbm5
	}
	\let\sl\it
	\twocolumn
	\sloppy
	\voffset=-2.54truecm
	\hoffset=-2.54truecm
	\flushbottom
	\parindent 1em
	\leftmargini 2em
	\leftmarginv .5em
	\leftmarginvi .5em
	\marginparwidth 48pt
	\marginparsep 10pt
	\setlength{\columnsep}{2truecm}
	\setlength{\textwidth}{25.4truecm}
	\setlength{\textheight}{17truecm}
>	\baselineskip=16pt
	\oddsidemargin .18truein
	\evensidemargin .17truein
}

\def\eqalign#1{\null\,\vcenter{\openup\jot\m@th
  \ialign{\strut\hfil$\displaystyle{##}$&$\displaystyle{{}##}$\hfil
      \crcr#1\crcr}}\,}
\def\eqalignno#1{\displ@y \tabskip\centering
  \halign
to\displaywidth{\hfil$\@lign\displaystyle{##}$\tabskip\z@skip
    &$\@lign\displaystyle{{}##}$\hfil\tabskip\centering
    &\llap{$\@lign##$}\tabskip\z@skip\crcr
    #1\crcr}}

\def\section{\@startsection {section}{1}{\z@}{3.ex plus 1ex minus
 .2ex}{2.ex plus .2ex}{\large\bf}}
\def\subsection{\@startsection{subsection}{2}{\z@}{2.75ex plus 1ex
minus
 .2ex}{1.5ex plus .2ex}{\bf}}

\def\appendix{{\newpage\section*{Appendix}}\let\appendix\section%
	{\setcounter{section}{0}
	\gdef\thesection{\Alph{section}}}\section}

\def\abstract{\if@twocolumn
\section*{Abstract}
\else 
\begin{center}
{\bf Abstract\vspace{-.5em}\vspace{0pt}}
\end{center}
\quotation
\fi}

\catcode`\@=12

\def\ibid{{\it ibid.\/}}

\newcommand{\beq}{\begin{equation}}
\newcommand{\eeq}{\end{equation}}
\newcommand{\beqa}{\begin{eqnarray}}
\newcommand{\eeqa}{\end{eqnarray}}

\def\noj#1,#2,{{\bf #1} (19#2)\ }
\def\jou#1,#2,#3,{{\sl #1\/ }{\bf #2} (19#3)\ }
\def\ann#1,#2,{{\sl Ann.\ Physics\/ }{\bf #1} (19#2)\ }
\def\cmp#1,#2,{{\sl Comm.\ Math.\ Phys.\/ }{\bf #1} (19#2)\ }
\def\ma#1,#2,{{\sl Math.\ Ann.\/ }{\bf #1} (19#2)\ }
\def\jd#1,#2,{{\sl J.\ Diff.\ Geom.\/ }{\bf #1} (19#2)\ }
\def\invm#1,#2,{{\sl Invent.\ Math.\/ }{\bf #1} (19#2)\ }
\def\cq#1,#2,{{\sl Class.\ Quantum Grav.\/ }{\bf #1} (19#2)\ }
\def\cqg#1,#2,{{\sl Class.\ Quantum Grav.\/ }{\bf #1} (19#2)\ }
\def\ijmp#1,#2,{{\sl Int.\ J.\ Mod.\ Phys.\/ }{\bf A#1} (19#2)\ }
\def\jmphy#1,#2,{{\sl J.\ Geom.\ Phys.\/ }{\bf #1} (19#2)\ }
\def\jams#1,#2,{{\sl J.\ Amer.\ Math.\ Soc.\/ }{\bf #1} (19#2)\ }
\def\grg#1,#2,{{\sl Gen.\ Rel.\ Grav.\/ }{\bf #1} (19#2)\ }
\def\mpl#1,#2,{{\sl Mod.\ Phys.\ Lett.\/ }{\bf A#1} (19#2)\ }
\def\nc#1,#2,{{\sl Nuovo Cim.\/ }{\bf #1} (19#2)\ }
\def\np#1,#2,{{\sl Nucl.\ Phys.\/ }{\bf B#1} (19#2)\ }
\def\pl#1,#2,{{\sl Phys.\ Lett.\/ }{\bf #1B} (19#2)\ }
\def\pla#1,#2,{{\sl Phys.\ Lett.\/ }{\bf #1A} (19#2)\ }
\def\pr#1,#2,{{\sl Phys.\ Rev.\/ }{\bf #1} (19#2)\ }
\def\prd#1,#2,{{\sl Phys.\ Rev.\/ }{\bf D#1} (19#2)\ }
\def\prl#1,#2,{{\sl Phys.\ Rev.\ Lett.\/ }{\bf #1} (19#2)\ }
\def\prp#1,#2,{{\sl Phys.\ Rept.\/ }{\bf #1C} (19#2)\ }
\def\ptp#1,#2,{{\sl Prog.\ Theor.\ Phys.\/ }{\bf #1} (19#2)\ }
\def\ptpsup#1,#2,{{\sl Prog.\ Theor.\ Phys.\/ Suppl.\/ }{\bf #1}
(19#2)\ }
\def\rmp#1,#2,{{\sl Rev.\ Mod.\ Phys.\/ }{\bf #1} (19#2)\ }
\def\yadfiz#1,#2,#3[#4,#5]{{\sl Yad.\ Fiz.\/ }{\bf #1} (19#2) #3%
\ [{\sl Sov.\ J.\ Nucl.\ Phys.\/ }{\bf #4} (19#2) #5]}
\def\zh#1,#2,#3[#4,#5]{{\sl Zh..\ Exp.\ Theor.\ Fiz.\/ }{\bf #1}
(19#2) #3%
\ [{\sl Sov.\ Phys.\ JETP\/ }{\bf #4} (19#2) #5]}

\def\beq{\begin{equation}}
\def\eeq{\end{equation}}
\def\beqar{\begin{eqnarray}}
\def\eeqar{\end{eqnarray}}

\def\nfrac#1#2{{\displaystyle{\vphantom1\smash{\lower.5ex\hbox{\sma
ll$#1$}}%
	\over\vphantom1\smash{\raise.25ex\hbox{\small$#2$}}}}}

\def\n#1{\mskip-#1mu}

\def\to{\rightarrow}

\def\lae{\mathrel{\mathop{\smash{\lower .5 ex
\hbox{$\stackrel<\sim$}}}}}
\def\lae{\mathrel{\mathop{\smash{\lower .5 ex
\hbox{$\stackrel>\sim$}}}}}

\def\l:{\mathopen{:}\,}
\def\r:{\,\mathclose{:}}

\catcode`\@=11
\def\theequation{\arabic{equation}}
\catcode`\@=12

\nblack
\bcites

\nblack

\catcode`\@=11
\def\theequation{\thesection.\arabic{equation}}
\@addtoreset{equation}{section}
\@addtoreset{footnote}{section}
\@addtoreset{footnote}{subsection}
\catcode`\@=12

\newcommand{\beqn}{\begin{equation}}
\newcommand{\eeqn}{\end{equation}}
\newcommand{\beqnarray}{\begin{eqnarray}}
\newcommand{\eeqnarray}{\end{eqnarray}}
\newcommand {\bear} [1] {\begin {array} {#1}}
\newcommand {\ear} {\end {array}}

\newcommand {\beqarn} {\begin{eqnarray*}}
\newcommand {\eeqarn} {\end{eqnarray*}}

\newcommand {\mrm} [1] {\mathrm {#1}}

\newcommand {\myref} [1]	%
	{%
	\begin{thebibliography} {99}	%
			{#1}	%
	\end {thebibliography}}

\def\emph#1{{\em #1}}
\def\mathbf#1{{\bf #1}}
\def\mathrm#1{{\rm #1}}
\def\mathit#1{{\it #1}}

\catcode`\@=11
\@addtoreset{footnote}{section}
\catcode`\@=12

\newcounter	{exinsert}	[subsection]

\renewcommand {\theexinsert}	%
{	\thesubsection.\arabic{equation}}

\renewcommand {\theequation} {\thesection.\arabic{equation}}

\catcode`\@=11
\@addtoreset{equation}{section}
\catcode`\@=12

\newenvironment {exinsert} [1]	%
{	
	\begin {quotation}	%
	\refstepcounter {equation}	%
	{\bf {#1}} \theequation. \,\,	%
}
{	
	\end {quotation}
}

\newcommand {\bprop}
{\begin {exinsert} {Proposition}
}
\newcommand {\eprop} {\end {exinsert} }

\newcommand {\bexe} {\begin {exinsert} {Exercise}}
\newcommand {\eexe} {\end {exinsert} }

\newcommand {\bexa} {\begin {exinsert} {Example}}
\newcommand {\eexa} {\end {exinsert} }

\newcommand {\literal} [1] {{ \mbox {$\mrm {#1}$}}}

\newcommand {\grad} {\nabla}

\newcommand {\ncmd} [2] {\newcommand {#1} {#2}}

\ncmd {\vgrad} {{\vec \grad}}

\ncmd {\Mv} {{\cal M}_V}
\ncmd {\Mh} {{\cal M}_H}

\ncmd {\adj} {\literal {adj}}	
\ncmd {\Mp} {M_\literal {planck}}	

\ncmd {\x} {$\times$}

\newcommand {\btab} [3]
  {\begin{table} [htb]
   \begin{center}
   \caption {#2}	\label {tab:#1}
   \begin {tabular} {#3}
   \hline}

\newcommand {\etab}
	{  \hline   \end {tabular}
   \end{center}
  \end{table}}

\ncmd {\bR} {\textbf {R}}
\ncmd {\bC} {\textbf {C}}

\ncmd {\CQ} {\cal Q}

\def\jou#1,#2,#3,{{\sl #1\/ }{\bf #2} (19#3)\ }
\def\ibid#1,#2,{{\sl ibid.\/ }{\bf #1} (19#2)\ }
\def\am#1,#2,{{\sl Acta. Math.\/ } {\bf #1} (19#2)\ }
\def\annp#1,#2,{{\sl Ann.\ Physics\/ }{\bf #1} (19#2)\ }
\def\cmp#1,#2,{{\sl Comm.\ Math.\ Phys.\/ }{\bf #1} (19#2)\ }
\def\cqg#1,#2,{{\sl Class.\ Quantum Grav.\/ }{\bf #1} (19#2)\ }
\def\dm#1,#2,{{\sl Duke\ Math.\ J.\/ }{\bf #1} (19#2)\ }
\def\ijmpa#1,#2,{{\sl Int.\ J.\ Mod.\ Phys.\/ }{\bf A#1} (19#2)\ }
\def\invm#1,#2,{{\sl Invent.\ Math.\/ }{\bf #1} (19#2)\ }
\def\jhep#1,#2,{{\sl JHEP\/ }{\bf #1} (19#2)\ }
\def\jmp#1,#2,{{\sl J.\ Geom.\ Phys.\/ }{\bf #1} (19#2)\ }
\def\jams#1,#2,{{\sl J.\ Diff.\ Geom.\/ }{\bf #1} (19#2)\ }
\def\jdg#1,#2,{{\sl J.\ Amer.\ Math.\ Soc.\/ }{\bf #1} (19#2)\ }
\def\jpa#1,#2,{{\sl J.\ Phys.\ A.\/ }{\bf #1} (19#2)\ }
\def\grg#1,#2,{{\sl Gen.\ Rel.\ Grav.\/ }{\bf #1} (19#2)\ }
\def\mpla#1,#2,{{\sl Mod.\ Phys.\ Lett.\/ }{\bf A#1} (19#2)\ }
\def\ma#1,#2,{{\sl Math.\ Ann.\/ } {\bf #1} (19#2)\ }
\def\nc#1,#2,{{\sl Nuovo\ Cim.\/ }{\bf #1} (19#2)\ }
\def\npb#1,#2,{{\sl Nucl.\ Phys.\/ }{\bf B#1} (19#2)\ }
\def\plb#1,#2,{{\sl Phys.\ Lett.\/ }{\bf #1B} (19#2)\ }
\def\pla#1,#2,{{\sl Phys.\ Lett.\/ }{\bf #1A} (19#2)\ }
\def\pnas#1,#2,{{\sl Proc..Nat.Acad.Sci.\/ }{\bf #1} (19#2)\ }
\def\prev#1,#2,{{\sl Phys.\ Rev.\/ }{\bf #1} (19#2)\ }
\def\prd#1,#2,{{\sl Phys.\ Rev.\/ }{\bf D#1} (19#2)\ }
\def\prl#1,#2,{{\sl Phys.\ Rev.\ Lett.\/ }{\bf #1} (19#2)\ }
\def\prpt#1,#2,{{\sl Phys.\ Rept.\/ }{\bf #1C} (19#2)\ }
\def\ptp#1,#2,{{\sl Prog.\ Theor.\ Phys.\/ }{\bf #1} (19#2)\ }
\def\ptpsup#1,#2,{{\sl Prog.\ Theor.\ Phys.\/ Suppl.\/ }{\bf #1} (19#2)\ }
\def\rmp#1,#2,{{\sl Rev.\ Mod.\ Phys.\/ }{\bf #1} (19#2)\ }
\def\sm#1,#2,{{\sl Selec. Math.\/ }{\bf #1} (19#2)\ }
\def\yadfiz#1,#2,#3[#4,#5]{{\sl Yad.\ Fiz.\/ }{\bf #1} (19#2) #3%
\ [{\sl Sov.\ J.\ Nucl.\ Phys.\/ }{\bf #4} (19#2) #5]}
\def\zh#1,#2,#3[#4,#5]{{\sl Zh.\ Exp.\ Theor.\ Fiz.\/ }{\bf #1} (19#2) #3%
\ [{\sl Sov.\ Phys.\ JETP\/ }{\bf #4} (19#2) #5]}

\begin {document}


\begin{titlepage}

\begin{center}
\hfill
UCB--PTH--98/49,~~
LBNL--42387,~~SPIN-1998/1 \\
\hfill hep-th/9810128

\vskip 1.5 cm
{\Large \bf On The Field Theory Limit Of D-Instantons}
\vskip 1 cm
{\large Hirosi Ooguri$^{1,2}$ and Kostas Skenderis$^3$}\\
\vskip 1cm
{$^1$ Department of Physics,
University of California at Berkeley,\\
Berkeley, California 94720}\\
\vskip .3in
{$^2$ Theoretical Physics Group, Mail Stop 50A-5101,\\
Lawrence Berkeley National Laboratory,\\
Berkeley, California 94720}\\
\vskip .3in
{$^3$ Spinoza Institute, University of Utrecht,\\
Leuvenlaan 4, 3584 CE Utrecht, The Netherlands}

\end{center}

\vskip 0.5 cm
\begin{abstract}

We study the dilaton/axion configuration near D-instantons
in type IIB superstring theory. In the field theory limit, 
the metric near the instantons becomes flat in the string frame 
as well as in the Einstein frame. In the large $N$ limit,
the string coupling constant becomes zero except near the origin.
The supersymmetry of this configuration is analyzed. 
An implication of this result to the IIB Matrix Model is discussed. 

\end{abstract}
\end{titlepage}

\section{Introduction}

According to the AdS/CFT correspondence \cite{mal,gkp,w}, 
string theory on
$(p+2)$-dimensional Anti-de Sitter Space ($AdS_{p+2}$) times a
compact space is equivalent to a $(p+1)$-dimensional conformal field
theory.
In particular, string theory on $AdS_4 \times S^7$, $AdS_5 \times S^5$
and $AdS_7 \times S^4$ are shown to be equivalent to conformal field theory
on M$2$, D$3$ and M$5$ branes, respectively. 
These correspondences were discovered by studying the near horizon region
of $p$-branes in two different ways. One is to consider strings propagating
in the curved background generated by the $p$-brane. Another is to use
the collective coordinates of the $p$-brane, which in some limit is
a $(p+1)$-dimensional conformal field theory. 
The equivalence of the two descriptions implies the correspondence. 

In this paper, we study the region near ($-1$)-branes, 
$i.e.$ instantons. This case is of
particular interest since the corresponding $0$-dimensional gauge theory
has been conjectured to give a non-perturbative definition of the type IIB
superstring \cite{ikkt,fkkt}. 
We study the dilaton/axion fields configuration 
near the instantons. If we take the field theory limit of Sen and
Seiberg \cite{sen,seiberg},
the metric near the instantons becomes flat in the string frame
as well as in the Einstein frame. Moreover, in the large $N$ limit,
the string coupling constant becomes zero except near the origin.
This seems in support of the conjecture \cite{ikkt,fkkt}. The
situation in the case of finite $N$ is subtle and we will
comment on this case toward the end of this paper.

\section{Wick Rotation and Supersymmetry}

The instanton solution is defined in the Euclidean signature
space. Since the Wick rotation of type IIB supergravity
is subtle, we  would like to start our discussion by
stating our prescriptions for the Wick rotation and
supersymmetry in the Euclidean signature space. 

In the Minkowski signature space, 
the dilaton and axion fields of type IIB theory parametrize 
the upper half-plane, or
the coset space $SL(2,R)/U(1)$. Following \cite{jhs}, 
we introduce the frame field,
\beq
  V = \left( \matrix{
      V^1_- & V^1_+ \cr V^2_- & V^2_+} \right),
\label{matV}
\eeq
and define the local $U(1)$ action as
\beq
 V \rightarrow V \left( \matrix{ e^{-i\Sigma} & 0 \cr
         0 & e^{i\Sigma} } \right)
\label{uone}
\eeq
where $\Sigma$ is a $U(1)$ phase. 
It is convenient to parametrize the matrix $V$ as
\beq
 V = \frac{1}{\sqrt{-2i\tau_2}}
 \left( \matrix{ \bar{\tau} e^{-i\lambda} & \tau e^{i\lambda} \cr
       e^{-i\lambda} & e^{i\lambda} } \right).
\label{param}
\eeq
We can fix the $U(1)$ gauge symmetry by setting
the scalar field $\lambda$ to be a function of $\tau$.
For example, $\lambda$ is set equal to 
$-{\sl Im} {\rm ~log}(\tau + i)$ in \cite{ggp} whereas
$\lambda = 0$ is used in \cite{gs}.

The complex scalar field $\tau$ in (\ref{param}) is related to the
dilaton $\phi$ and the axion $a$ as
\beq
  \tau = a + i e^{-\phi}.
\eeq
To write the type IIB supergravity equations of motion, we introduce
two $SL(2,R)$ singlet currents,
\beqa
  P_\mu & = & - \epsilon_{\alpha\beta} V_+^\alpha \partial_\mu
 V_+^\beta
= \frac{i}{2} \frac{\partial_\mu \tau}{\tau_2} e^{2i\lambda} \nonumber \\
Q_\mu & = &
-i \epsilon_{\alpha\beta} V_-^\alpha \partial_\mu V_+^\beta
 = \partial_\mu \lambda - \frac{1}{2} \frac{\partial_\mu \tau_1}{\tau_2}.
\label{currents}
\eeqa
Under the $U(1)$ gauge symmetry (\ref{uone}), they transform
as
\beqa
  P_\mu & \rightarrow & P_\mu e^{2i\Sigma} \nonumber \\
  Q_\mu & \rightarrow & Q_\mu + \partial_\mu \Sigma. 
\eeqa
The equations of motion (in the absence of the $p$-form fields,
$p=2,4,6$) are
\beqa
 R_{\mu\nu} &=& P_\mu P^*_\nu + P^*_\mu P_\nu \nonumber \\
 D^\mu P_\mu &=&  (\nabla_\mu - 2i Q_\mu) P_\mu = 0,
\eeqa
where $R_{\mu\nu}$ is the Ricci tensor. 
Substituting (\ref{currents}) into these, the equations of motion
can be expressed in terms of $\phi$ and $a$ as
\beqa
&& R_{\mu\nu} = \frac{1}{2} (\partial_\mu \phi \partial_\nu \phi
+ e^{2\phi} \partial_\mu a \partial_\nu a) 
\nonumber \\
&&  \Delta a + 2 \partial^\mu \phi \partial_\mu a = 0 
\nonumber \\
&&  \Delta \phi - e^{2\phi} (\partial a)^2  = 0.
\label{mink}
\eeqa
These can be derived from the Lagrangian density
\beq
 {\cal L} = R - \frac{1}{2} (\partial \phi)^2
              - \frac{1}{2} e^{2\phi} (\partial a)^2.
\eeq
The supersymmetry transformations of the dilatino $\rho$ and
the gravitino $\psi_\mu$ are given by
\beqa
  \delta \rho &=& i P_\mu \gamma^\mu \epsilon^* \nonumber \\
  \delta \psi_\mu &=& \left( \nabla_\mu - \frac{i}{2} Q_\mu
\right) \epsilon .
\label{susy}
\eeqa

The instanton is a solution in Euclidean signature space. The 
equations is this case are obtained from (\ref{mink}) by the 
substitution $a \to \alpha = i a$,
\beqa
&& R_{\mu\nu} = \frac{1}{2} (\partial_\mu \phi \partial_\nu \phi
- e^{2\phi} \partial_\mu \alpha \partial_\nu \alpha) 
\nonumber \\
&&  \Delta \alpha + 2 \partial^\mu \phi \partial_\mu \alpha = 0 
\nonumber \\
&&  \Delta \phi + e^{2\phi} (\partial \alpha)^2  = 0.
\label{eucl}
\eeqa
The supersymmetry transformation rules that can be derived 
analogously \cite{nicolai,pvn,wald}.
We make the substitution $a \to \alpha = i a$, and we treat 
$\epsilon$ and $\epsilon^*$ as independent spinors.
In addition, we Wick rotate the spinors as in \cite{pvn,wald}. 
This yields
\beqa
&&\delta \rho = i P_\mu \gamma^\mu \epsilon^*, \qquad 
\delta \rho^* = - i P^*_\mu \gamma^\mu \epsilon  \nonumber \\
&&\delta \psi^\mu = (\nabla_\mu - {i \over 2} Q_\mu) \epsilon, \qquad
\delta \psi^{\mu *} = (\nabla_\mu + {i \over 2} Q_\mu) \epsilon^*,
\label{euclidsusy}
\eeqa
where
\beqa
&& P_\mu = {1 \over 2} {(\partial_\mu \tau_1 - \partial_\mu \tau_2) 
\over \tau_2}
e^{2i\lambda}, \nonumber \\
&&P_\mu^* = - {1 \over 2} {(\partial_\mu \tau_1 + \partial_\mu \tau_2) 
\over \tau_2}
e^{-2i\lambda}, \qquad
\nonumber \\
&&Q_\mu = \partial_\mu \lambda + {i \over 2} 
{\partial_\mu \tau_1 \over \tau_2}.
\label{Qwick}
\eeqa
The invariance of the Wick-rotated action under these rules directly follows
from the invariance of the original action under the transformation 
rules (\ref{susy}). For a more complete discussion of 
Euclidean spinors and Wick rotation we refer to \cite{nicolai,pvn,wald}.

\section{Instanton Solution}

As shown in \cite{ggp}, the Euclidean equations of motion
 (\ref{eucl}) have a solution
where the metric (in the Einstein frame)
is flat $g_{\mu\nu} = \delta_{\mu\nu}$
and the dilaton and the axion are related as
\beq
   \partial_\mu \alpha =  \pm e^{-\phi} \partial_\mu \phi.
\label{axion}
\eeq
In the following, we choose the plus sign in the right hand side.
The equations (\ref{eucl}) are then satisfied if the dilaton obeys
\beq
  \Delta e^\phi = 0.
\label{dilaton}
\eeq

A spherically symmetric solution to (\ref{dilaton}) with
the boundary condition $e^\phi \rightarrow g_s$ at infinity 
$r= \infty$ is given by
\beq
     e^\phi = g_s \left( 1 + \frac{c}{r^8} \right)
\eeq
for some constant $c$. The equation (\ref{axion}) then determines
the axion $\alpha$ as
\beq
    \alpha = \frac{-1}{g_s \left( 1 + \frac{c}{r^8} \right)}
       + {\rm const}.
\eeq
Since $\alpha$ behaves for large $r$ as
\beq
   \alpha \simeq \frac{c}{g_s} \frac{1}{r^8} + \cdots,
\eeq
the constant $c$ is related to the instanton
charge $N$ as 
\beq
    c = c_0 g_s N l_s^8
\eeq
where $l_s$ is the string length and
$c_0$ is a numerical constant related to the volume of
the unit 9-sphere.

To summarize, the instanton solution with $N$ unites of
charge is given by
\beqa
 && g_{\mu\nu} = \delta_{\mu\nu} \nonumber \\
  && e^\phi = g_s \left( 1 + c_0 \frac{g_s N l_s^8}{r^8} \right)
 \nonumber \\
  &&\alpha =  - g_s^{-1} 
 \left( 1 + c_0\frac{g_s N l_s^8}{r^8} \right)^{-1}
       + {\rm const}.
\label{solution}
\eeqa

\section{Field Theory Limit of D-Instantons}

It was shown in \cite{ggp} that the solution 
(\ref{solution}) preserves half of the maximal supersymmetry. 
Here we will study the field theory limit \cite{mal,sen,seiberg}
\beq
  u = \frac{r}{l_s^2}, ~~g_{YM}^2 = \frac{g_s}{l_s^4} : {\rm fixed}, 
~~~~ l_s \rightarrow 0,
\label{dkps}
\eeq
of the solutions\footnote{
The near instanton configuration 
was also studied in \cite{BB} (see also \cite{KK} for a related
observation). In that paper they considered
the strict $r \to 0$ limit instead of the field
theory limit (\ref{dkps}). They showed that 
the configuration one obtains in this limit preserves the maximal
supersymmetry 
(a similar result holds for all D$p$-branes ($p<7$) \cite{BPS}).
Notice that the final configuration they obtain is singular 
as the dilaton diverges in the limit. In the field theory limit
(\ref{dkps}), the dilaton and axion remain finite as in 
(\ref{neardilaton})}.
In this limit, the $l_s$-dependence of the dilaton/axion fields 
in (\ref{solution}) disappears and we obtain
\beq
   \tau_2 = e^{-\phi} \simeq \frac{Nu^8}{c_0(g_{YM}^2 N)^2},
~~\tau_1 = - \tau_2 + {\rm const}.
\label{neardilaton}
\eeq
Since $\partial_\mu \tau_1 = -\partial_\mu \tau_2$ for the 
instanton solution (\ref{solution}), the composite gauge field 
$Q_{\mu}$ given by (\ref{Qwick}) is a pure gauge\footnote{This is
the case even before the field theory limit (\ref{dkps})
is taken.}. Therefore
it is natural to set $Q_\mu=0$ by fixing the $U(1)$ gauge symmetry as
\beq
 \lambda = \frac{i}{2} {\rm log} \tau_2
\label{gaugecond}
\eeq
In this gauge, the gravitino variation in (\ref{euclidsusy})  
becomes simply
\beq
  \delta \psi_\mu = \partial_\mu \epsilon.
\eeq
Thus the equation $\delta \psi_\mu = 0$ has the maximal number 
of solutions. 

Let us turn to the dilatino variation in (\ref{euclidsusy}). 
Since $\partial_\mu \tau_1 = - \partial_\mu \tau_2$, $P^*_\mu=0$
and therefore, $\delta \rho^*=0$. On the other hand,
\beq
P_\mu = \partial_\mu {1 \over \tau_2} = 
\frac{c_0(g_{YM}^2 N)^2}{N} \partial_\mu u^{-8}
\label{breaking}
\eeq
which is non-zero in the field theory limit. Thus, 
only $1/2$ of supersymmetry is preserved even in the
field theory limit.

\section{Comment on the IIB Matrix Model}

In the D-brane description of the $p$-brane \cite{pol},
the open string dynamics on the brane reduces to the $(p+1)$-dimensional
supersymmetric gauge theory \cite{gauge} in the limit 
\beq
   u = \frac{r}{l_s^2},~~g_{YM}^2 = \frac{g_s}{l_s^{3-p}}:
{\rm fixed},~~~~~~l_s \rightarrow 0.
\eeq
Repeating
the argument in \cite{mal} in the case of the D($-1$) brane, 
it is natural to expect that
type IIB string in the flat metric (\ref{stringmetric}) and
the dilaton background (\ref{neardilaton}) is equivalent to
the $0$-dimensional matrix model given by the action
\beq
 S = - \frac{1}{g_{YM}^2} {\sl tr}\left( \frac{1}{4} [A_\mu, A_\nu]
[A^\mu, A^\nu] + \frac{1}{2} \bar{\psi} \Gamma^\mu
[A_\mu, \psi] \right),
\label{matrix}
\eeq
where $A_\mu$ ($\mu=1,...,10$) and $\psi$ are $N \times N$
hermitian matrices. 
The large $N$ limit of this model has been proposed in \cite{ikkt,fkkt}
as a non-perturbative definition of type IIB string theory.
There they found that the string length of the type IIB string
is $(g_{YM}^2 N)^{1/4}$ and that the string coupling constant
is $(N\epsilon^2)^{-1}$, where $\epsilon$ is a cut-off parameter
in the matrix integral. 

Let us compare their results with the field theory
limit of the instanton studied in this paper. 
Since the metric for the instanton solution
is flat in the Einstein frame,
the string frame metric for the instanton solution is given by
\beq
  ds_{s}^2 = l_s^4\sqrt{1+ c_0 \frac{g_{YM}^2 N}{l_s^4u^8}}
             (du^2 + u^2 d\Omega_9^2).
\eeq
Here $d\Omega_9$ is the line element of the unit 9-sphere.
In the limit (\ref{dkps}), this becomes
\beq
  \frac{ds_s^2}{l_s^2} \simeq
   \sqrt{c_0g_{YM}^2N}\left( \frac{du^2}{u^4} + \frac{d\Omega_9^2}{u^2}
                         \right)
     = \sqrt{c_0 g_{YM}^2 N} (d\tilde{u}^2 + \tilde{u}^2 d\Omega_9^2),
\label{stringmetric}
\eeq
where $\tilde{u} = 1/u$. Thus the metric in the string frame also
becomes flat. Moreover the factor $\sqrt{g_{YM}^2 N}$
is reminiscent of the string length found in \cite{fkkt}. 

At the same time, the dilaton in the field theory limit is
\beq
   e^\phi \simeq c_0 \frac{(g_{YM}^2 N)^2}{Nu^8}.
\label{ncd}
\eeq
This appears to be different from the expression for the string
coupling constant found in \cite{fkkt}. The non-constant
dilaton (\ref{ncd}) is also responsible for the breaking of
1/2 of supersymmetry near the instanton, as we saw in
(\ref{breaking}). On the other hand,
it was pointed out in \cite{ikkt} that the matrix model
(\ref{matrix}) has $N=2$ super Poincar\'{e} symmetry in ten dimensions.

In the limit $N \rightarrow \infty$ with $g_{YM}^2 N$ finite, 
these two view points are in complete agreement. 
In the field theory limit, the string coupling given by
the dilaton field $e^{\phi}$ is small for $u^8 \gg 1/N$. 
As we take $N$ to be large, the size of this region expands.
In the limit $N \rightarrow \infty$, one obtains type IIB string
in the flat space (\ref{stringmetric}) with vanishing string
coupling $e^\phi = 0$. From the large $N$ analysis of (\ref{matrix}),
one also finds free type IIB strings \cite{ikkt}. It would
be very interesting to clarify the situation at finite $N$. 

\section*{Acknowledgments}

We would like to thank Michael Green for discussions.
We also thank Tom Banks and Juan Maldacena for comments
on the earlier version of this paper.  
We thank the theory division of CERN where this work was
initiated for the hospitality.
K.S. would also like to thank 
the Aspen Center for Physics and the Erwin Schr\"{o}dinger 
Institute in Vienna where part of this work was completed 
for their hospitality.
The work of H.O. was supported in part by the NSF
grant PHY-95-14797 and the DOE grant DE-AC03-76SF00098.
The research of K.S. is supported by the Netherlands Organization 
for Scientific Research (NWO).

\end{document}